\documentclass[prb,aps,showpacs,twocolumn,preprintnumbers,amsmath,amssymb,superscriptaddress]{revtex4-1}
\usepackage{amsmath,amssymb,amsfonts}
\usepackage{graphicx}
\usepackage[colorlinks=True,linkcolor=red,citecolor=blue,urlcolor=blue]{hyperref}
\usepackage{xcolor}
\usepackage{chngcntr}
\usepackage{braket}
\usepackage{hyperref}
\usepackage{nicefrac}
\usepackage{bm}
\usepackage{bbm}
\usepackage{braket}

\renewcommand{\dag}{^{\dagger}}

\def\cK{\mathcal{K}}
\def\cU{\mathcal{U}}
\def\cA{\mathcal{A}}
\def\id{\mathbbm{1}}
\DeclareMathOperator{\pf}{pf}

\renewcommand{\dag}{^{\dagger}}

\begin{document}
\bibliographystyle{apsrev4-1}
\title{Generalizations of the Pfaffian to non-antisymmetric matrices}

\author{D\'{a}niel Varjas}
\affiliation{QuTech and Kavli Institute of Nanoscience, Delft University of Technology, P.O. Box 4056, 2600 GA Delft, The Netherlands}
\affiliation{Department of Physics, Stockholm University, AlbaNova University Center, 106 91 Stockholm, Sweden}
\email{dvarjas@gmail.com}

\date{\today}
\begin{abstract}
We study two generalizations of the Pfaffian to non-antisymmetric matrices and derive their properties and relation to each other.
The first approach is based on the Wigner normal-form, applicable to conjugate-normal matrices, and retains most properties of the Pfaffian, including that it is the square-root of the determinant.
The second approach is to take the Pfaffian of the antisymmetrized matrix, applicable to all matrices.
We show that this formulation is equivalent to substituting a non-antisymmetric matrix into the polynomial definition of the Pfaffian.
We find that the two definitions differ in a positive real factor, making the second definition violate the determinant identity.
\end{abstract}
\maketitle

\section{Introduction}
The determinant of a skew-symmetric matrix is a complete square when considered as a polynomial of the matrix entries.
The Pfaffian was first defined based on this observation, as the square root of this polynomial.
For real skew-symmetric matrices the Pfaffian is invariant under proper orthogonal transformations.
In quantum mechanics, the Pfaffian of complex matrices proved to be a valuable tool when characterizing antilinear operators.

Antiunitary operators play an important role in physics, as time-reversal and particle-hole symmetries are represented by antiunitary operators on the Hilbert space~\cite{wigner1932}.
Such operators have the form $\cU = U \cK$ where $U$ is unitary and $\cK$ is the complex conjugation operator.
The square of an antiunitary $\cU^2 = U U^*$ is unitary, and in the case of time-reversal and particle-hole symmetries $\cU^2 = \pm \id$.
The case of $\cU^2 = - \id$, relevant to time-reversal of half-integer spin fermions, implies $U = -U^T$, allowing the use of the Pfaffian to analyze representations of the antiunitary symmetry and the associated topological invariants~\cite{FuKane2006, Ryu2010}.

More generally, however, antiunitary symmetries which do not square to $\pm\id$ also occur, for example in magnetic symmetry groups.
For example, a magnetically ordered structure may be invariant under fourfold rotation combined with time-reversal (which reverses magnetization), but not the two operations separately.~\cite{Alexandradinata2016,Li2020, Araya2022}
The square of this operator is a twofold rotation, which is generally not proportional to the identity.

In this note we extend the formalism offered by the Pfaffian to such cases.
First we review the Wigner normal form of normal antilinear operators, and define the generalized Pfaffian for these.
As we show, this definition maintains most of the algebraic properties of the Pfaffian.
Next, we consider an alternative definition by taking the Pfaffian of the antisymmetrized matrix.
This definition has the advantage of being a polynomial of the matrix entries, and only differs from the previous definition in a positive real factor, however, it is no longer the square root of the determinant.
Finally, we provide an algorithm to compute the normal form and the generalized Pfaffian.

\section{Normal form of antilinear operators}
We rely on the result of Wigner~\cite{Wigner1960} showing that all antiunitary operators admit a normal form, and its extension by Herbut and Vujičić~\cite{Vujicic1967} to normal antilinear operators.
We reproduce the key steps here in a slightly different convention, readers familiar with the subject may skip to the end of this section for the definition of the normal form.

Let $\cA$ be an antilinear operator which always factorizes as $\cA = A \cK$ with $A$ a linear operator and $\cK$ the complex conjugation operator (a fixed antilinear involution with $\cK^2 = \id$).
It acts on a Hilbert-space $\mathcal{H}$ that is equipped with inner product $\braket{.,.}$, defining Hermitian adjoint through $\braket{v, \cA w} = \braket{\cA\dag v, w}^*$ for all $v, w \in \mathcal{H}$.
In the following we assume that $\cA$ is a normal antilinear operator, \emph{i.e.} it commutes with its adjoint: $\cA\dag\cA = \cA\cA\dag$; or equivalently $A$ is conjugate-normal: $A^T A^* = A A\dag$.
This condition is automatically satisfied for the common cases of symmetric or antisymmetric $A$ (hermitian or antihermitian $\cA$), as well as for unitary $A$ (antiunitary $\cA$).
Under unitary basis transformations of the Hilbert-space antilinear operators transform as
\begin{align}
\cA' &= U\dag \cA U \\
\intertext{or equivalently}
A' &= U\dag A U^*
\end{align}
where the columns of $U$ form a new orthonormal basis.
We show that there exists a basis where $A$ takes a particularly simple form, allowing the decomposition
\begin{equation}
\label{eqn:normal_form_2}
A = \tilde{U} \tilde{\Sigma} \tilde{U}^T
\end{equation}
where $\tilde{\Sigma}$ is hermitian and block-diagonal with $1\times 1$ positive real and $2\times 2$ off-diagonal blocks.

First we show that the eigenvalues of the normal linear operator $\Lambda = \cA^2 = A A^*$ are either real or come in complex conjugate pairs.
Let $\omega$ and $v$ be an eigenpair of $\Lambda$:
\begin{equation}
\Lambda v = \omega v.
\end{equation}
Then $w = \cA v = A v^*$ is an eigenvector with eigenvalue $\omega^*$:
\begin{equation}
\label{eqn:lambda_spaces}
\Lambda w = \cA^3 v = \cA \left(\Lambda v\right) = \cA \left(\omega v\right) = \omega^* \cA v = \omega^* w.
\end{equation}
As $\Lambda$ is normal, its eigenvectors belonging to different eigenvalues are orthogonal.
This shows that $\cA$ only mixes eigensubspaces of $\Lambda$ belonging to a pair of complex conjugate eigenvalues.

For non-negative real eigenvalues $\omega \geq 0$  it is possible to choose vectors that are invariant under $\cA$ and the corresponding block in the normal form is $1\times 1$ containing $\sqrt{\omega} \geq 0$~\cite{wigner1932,Vujicic1967}.

Next we show that $\cA$ can be brought to an off-diagonal form if all eigenvalues of $\Lambda$ are complex or negative real.
In the well known case of antiunitary $\cA$ with $\cA^2 = -\id$, it can be shown that $v$ and $\cA v$ are always orthogonal for arbitrary choice of $v$ following the argument used to prove Kramers' degeneracy~\cite{kramers1930,wigner1932}:
\begin{equation}
\braket{v, \cA v} = \braket{\cA v, \cA^2 v}^* = -\braket{\cA v, v}^* = -\braket{v, \cA v} = 0.
\end{equation}
In the following we use a similar argument for the case of normal $\cA$.
If $\cA$ and $\cA\dag$ commute, $M = \cA\dag\cA$ and $\Lambda = \cA^2$ also commute and can be simultaneously diagonalized.
Let $v$ be a common eigenvector such that $\Lambda v = \omega v$ with $\omega \ngeq 0$ and $M v = \mu v$ with $\mu \geq 0$.
The last condition follows from $M$ being positive semidefinite, and as $\Lambda\Lambda\dag = M^2$, the eigenvalues are related as $|\omega| = \mu$.
Writing the overlap as
\begin{align}
\braket{\cA v, v} =& \frac{1}{\omega} \braket{\cA v, \cA^2 v} = \frac{1}{\omega} \braket{\cA\dag \cA v, \cA v}^* \\
 =& \frac{1}{\omega} \braket{\cA v, \cA\dag \cA v} = \frac{\mu}{\omega} \braket{\cA v, v} = 0,
\end{align}
which vanishes because $\mu \neq \omega$, showing that $v$ and $\cA v$ are orthogonal.
This can also be interpreted as the diagonal matrix elements of $\cA$ vanishing.

Let $v$ be normalized such that $\braket{v, v} = 1$, and choose
\begin{equation}
w = \frac{\sqrt{\omega}}{\mu} \cA v,
\end{equation}
this guarantees that $w$ is also normalized. 
For the off-diagonal matrix elements of $\cA$ we find
\begin{align}
\braket{v, \cA w} =& \frac{\sqrt{\omega}^*}{\mu} \braket{v, \cA^2 v} = \frac{\omega \sqrt{\omega}^*}{\mu} = \sqrt{\omega}\\
\braket{w, \cA v} =& \frac{\sqrt{\omega}^*}{\mu} \braket{\cA v, \cA v} =  \frac{\sqrt{\omega}^*}{\mu} \braket{v, \cA\dag \cA v} = \sqrt{\omega}^* .
\end{align}
As $v$ is an eigenvector of $\cA^2$, the subspace spanned by $v$ and $w$ is closed under the action of $\cA$, all matrix elements with states orthogonal to this subspace vanish.

Hence for a pair of complex eigenvalues of $\Lambda$ it is always possible to choose a basis where the normal form of $A$ consists of $2\times2$ off-diagonal blocks of the form
\begin{equation}
\left(\begin{array}{cc}
0 & \sqrt{\omega} \\
\sqrt{\omega}^* & 0
\end{array}\right)
\end{equation}
where $\operatorname{Im} \omega > 0$ and the branch cut in the square root is chosen at the negative real axis.


In the case of $\omega < 0$, we can restrict $\cA$ to this eigensubspace where $\Lambda = \omega\id$.
Now $\sqrt{\omega}$ is purely imaginary, and as seen from the above off-diagonal matrix elements, the restriction of $\Lambda$ to this subspace is antisymmetric.
Hence the known methods to obtain the normal form and Pfaffian are applicable~\cite{}.
We note that there is a freedom in choosing the relative phases of the off-diagonal elements in the normal form with various conventions used in the literature, with our definition the blocks are purely imaginary, not purely real for antisymmetric $A$.

In summary, if $\cA = A \cK$ is normal, there exists a unitary basis transformation $U$ that brings $A$ to its normal form, such that $A = U \Sigma U^T$.
The normal form is given by
\begin{equation}
\label{eqn:normal_form}
\Sigma = U \dag A U^* =
\left(\begin{array}{cc}
0 & \sqrt{\Omega} \\
\sqrt{\Omega}^* & 0
\end{array}\right) \oplus \left[\bigoplus_{\omega_n \in \mathbb{R}^+_0}
\left(\omega_n \right)\right]
\end{equation}
where $\sqrt{\Omega}$ is a diagonal matrix composed of the square roots of half of the complex or negative real eigenvalues of $\Lambda$ with $\operatorname{Im} \omega_n \geq 0$.
We take the square root with the branch cut along the negative real axis, and only include half of the negative real $\omega_n$ in the matrix $\Omega$.
This normal form differs in the ordering of the basis vectors from \eqref{eqn:normal_form_2}, but is identical otherwise.
\footnote{{
In the literature an alternate ordering of the basis for the normal form $A = \tilde{U}\tilde{\Sigma} \tilde{U}^T$ is frequently used with
\begin{equation*}
\tilde{\Sigma} =
\left[\bigoplus_{\omega_n \notin \mathbb{R}^+_0, \operatorname{Im} \omega_n \geq 0}
\left(\begin{array}{cc}
0 & \sqrt{\omega_n} \\
\sqrt{\omega_n}^* & 0
\end{array}\right)\right] \oplus \left[\bigoplus_{\omega_n \in \mathbb{R}^+_0}
\left(\omega_n \right)\right].
\end{equation*}
In this convention the Pfaffian is given by $\pf(A) = i^{n} \det (\tilde{U}) \sqrt{|\det(\tilde{\Sigma})|}$.
}}

If the spectrum of $\Lambda$ does not include the non-negative real axis, only the first term appears.
This statement is equivalent to the antisymmetrized matrix $\frac{1}{2}\left(A - A^T\right)$ being nonsingular for conjugate-normal $A$.
The basis where $A$ takes its normal form, also brings its antisymmetrized to its normal form:
\begin{equation}
\label{eqn:as_nf}
\frac{A - A^T}{2} = U \frac{\Sigma - \Sigma^T}{2} U^T = U (i \operatorname{Im} \Sigma) U^T.
\end{equation}
This is singular if and only if some eigenvalue $\omega_n$ is non-negative real.

\section{Definition of the generalized Pfaffian}
With the normal form at hand, we define the generalized Pfaffian of a $2n\times 2n$ conjugate-normal matrix $A$ with non-singular antisymmetric part as
\begin{equation}
\label{eqn:gen_pf_def}
\pf(A) = i^{n^2} \det (U) \sqrt{|\det(\Sigma)|}
\end{equation}
where the normal form is given by $A = U \Sigma U^T$ as in equation \eqref{eqn:normal_form}.
We may extend the definition to singular matrices with $\det(A) = 0$ by setting $\pf(A) = 0$.

In the case of antisymmetric $A$ this coincides with the usual definition of the Pfaffian, as in this case the normal form is
\begin{equation}
\Sigma = \left(-\sigma_y\right) \otimes \sqrt{|\Omega|},
\end{equation}
where $\sigma_y$ is the $y$ Pauli matrix, with Pfaffian~\cite{Lomont1985}
\begin{equation}
\pf \left[\left(-\sigma_y\right) \otimes \sqrt{|\Omega|}\right] = i^{n^2} \det \left(\sqrt{|\Omega|} \right) = i^{n^2} \sqrt{|\det(\Sigma)|},
\end{equation}
and we use the identity
\begin{equation}
\label{eqn:trf_identity}
\pf\left(U A U^T\right) = \det(U) \pf(A).
\end{equation}

We show that the definition \eqref{eqn:gen_pf_def} is unambiguous because the freedom in the choice of $U$ that produces the normal form leaves $\det(U)$ invariant.
First we consider the ambiguity in the ordering of the eigenvalues in the diagonal matrix $\sqrt{\Omega}$.
Exchange of two different diagonal elements in $\sqrt{\Omega}$ (and $\sqrt{\Omega}^*$) is compensated by two column exchanges in $U$, hence no change in its determinant.
This shows that every normal form, regardless of the ordering of the eigenvalues, results in the same Pfaffian.

Next we consider the ambiguity in choosing $U$ when some eigenvalues are degenerate.
In the general case, in the subspaces with complex $\omega_n$ the off-diagonal entries in the normal form connect different eigensubspaces of $\Lambda$ with $\omega_n \neq \omega_n^*$ eigenvalues.
Any basis transformation leaving the normal form invariant must not mix different subspaces.
Consider a restriction of $\Lambda$ to a pair of possibly degenerate subspaces such that
\begin{equation}
\Lambda_n = \left(\begin{array}{cc}
\omega_n \id & 0 \\
0 & \omega_n^* \id
\end{array}\right).
\end{equation}
In this basis the restricted normal form of the restricted $A_n$ is
\begin{equation}
\Sigma_n = \left(\begin{array}{cc}
0 & \sqrt{\omega_n}\id \\
\sqrt{\omega_n}^*\id & 0
\end{array}\right).
\end{equation}
The most general unitary transformation leaving both $\Lambda_n$ and $\Sigma_n$ invariant has the form
\begin{equation}
\label{eqn:block_diag_U}
U_n = \left(\begin{array}{cc}
V & 0 \\
0 & V^*
\end{array}\right)
\end{equation}
with unitary $V$.
The determinant of such a transformation is $\det(U_n) = \det(V) \det(V)^* = 1$.

Restricted to a subspace with $\omega_n < 0$ the normal form of $A_n = - A_n^T$ is $\left(-\sigma_y\right) \otimes \sqrt{|\omega_n|}$ and any transformation $U$ leaving it invariant is symplectic and has $\det(U_n) = 1$, this can be seen by applying \eqref{eqn:trf_identity}.

\section{Properties and identities}

A simultaneous interchange of two rows and the corresponding columns changes the sign of the Pfaffian because the corresponding permutation matrix has determinant $-1$ and can be included in $U$ without changing $\Sigma$.
However, in general simultaneous row and column operations (multiplication with a constant or adding a row/column to another) do not preserve the normality of $\cA$, hence their effect on the Pfaffian is ill-defined.
This is related to the fact that the identity
\begin{equation}
\pf(U A U^T) = \det(U) \pf(A)
\end{equation}
applies by construction for unitary $U$, but not for a general $U$, as $U A U^T$ is in general not conjugate-normal.
If we restrict to matrices that are proportional to their transpose, $A = \alpha A^T$ with some complex $\alpha\ngeq 0$, the identity becomes valid for all $U$, which is seen using the determinant identity \eqref{eqn:pf_square} and the continuity of $\pf$.

The following identities involving the Pfaffian are also true for our generalization with conjugate-normal $A$ if $A A^*$ has no positive real eigenvalues:
\begin{equation}
\label{eqn:pf_square}
\pf(A)^2 = \det(A).
\end{equation}
The proof involves simply substituting the normal form and the definition of $\pf(A)$ and using that $\det(\Sigma) = (-1)^n |\det(\Sigma)|$.

For any $\lambda\in\mathbbm{C}$
\begin{equation}
\pf(\lambda A) = \lambda^n \pf(A).
\end{equation}
The identity is true for $\lambda = 0$, and if $\lambda\neq 0$ the normal form of $A' = \lambda A$ is related to that of $A$ through $\Sigma' = |\lambda| \Sigma$ and $U' = (\lambda/|\lambda|)^{1/2} U$.

\begin{equation}
\pf(A^T) = (-1)^n \pf(A).
\end{equation}
We use that the normal form of $A' = A^T$ is related to that of $A$ through $\Sigma' = \Sigma$, $U' = U (\sigma_x \otimes \id_{n})$.
The new term appearing in $U'$ is a permutation matrix such that $(\sigma_x \otimes \id_{n}) \Sigma (\sigma_x \otimes \id_{n}) = \Sigma^T$, and its determinant is $(-1)^n$.

\begin{equation}
\pf(A\dag) = \pf(A)^*.
\end{equation}
By taking the hermitian adjoint of the normal form of $A$ we find that the normal form of $A' = A\dag$ is $\Sigma' = \Sigma$, $U' = U^*$.

For non-singular $A$
\begin{equation}
\label{eqn:pf_inv}
\pf(A^{-1}) = \pf(A)^{-1}.
\end{equation}
Taking the inverse of the normal form of $A$ we get $A^{-1} = U^* \Sigma^{-1} U\dag$. This is a suitable normal form for $A^{-1}$, and we use that $\det (U^*) = \det (U)^{-1}$.

For $A_1$ and $A_2$ satisfying the same conditions as $A$:
\begin{equation}
\pf(A_1 \oplus A_2) = \pf(A_1) \pf(A_2),
\end{equation}
which follows because a block-diagonal transformation matrix brings a block-diagonal matrix to its normal form up to row and column interchanges.

Let $A$ be as above and $B = B^T$ symmetric $m\times m$ matrix, then for the tensor product
\begin{equation}
\pf(A \otimes B) = (-1)^{nm(m-1)/2} \pf(A)^m \det(B)^n.
\end{equation}
The proof follows a similar reasoning to the proof for antisymmetric $A$, see Ref.~\onlinecite{Lomont1985} for example.
We write both matrices in the normal form as $A = U_A \Sigma_A U_A^T$ and $B = U_B \Sigma_B U_B^T$ where $\Sigma_B$ is non-negative real diagonal.
The normal form of the tensor product is $A\otimes B = (U_A\otimes U_B) (\Sigma_A\otimes \Sigma_B) (U_A\otimes U_B)^T$, this is guaranteed to have the correct form as $\Sigma_B \geq 0$.
Writing the definition of the Pfaffian, using that $\det(U_A\otimes U_B) = \det(U_A)^m \det(U_B)^{2n}$ (and similarly for $\Sigma$) and collecting the powers of $i$ produces the result.

If $A$ is non-singular
\begin{equation}
\frac{\partial}{\partial x} \pf(A) = \frac{1}{2} \pf(A) \operatorname{tr}\left(A^{-1} \frac{\partial A}{\partial x}\right).
\label{eqn:pf_diff}
\end{equation}
This can be seen either by using $\pf(A) = \pm \sqrt{\det(A)}$ or by substituting the normal form, and using Jacobi's identity for the derivative of the determinant.

While this last equation implies that the generalized Pfaffian is infinitely differentiable around any $A$ with with a non-singular antisymmetric part, it is not possible to analytically continue the definition to the case when the antisymmetric part becomes singular.
To see this consider the matrix
\begin{equation}
A =
\left(\begin{array}{cc}
0 & 1 + i x \\
1 - ix & 0
\end{array}\right).
\end{equation}
For $x>0$ this matrix is in its normal form and $\pf(A) = |1 + i x|$.
For $x<0$ however, the normal form is given by exchanging the rows and columns, and $\pf(A) = -|1 + i x|$, showing the jump discontinuity in the Pfaffian at $x=0$.

\section{Antisymmetrized Pfaffian}
Motivated by Ref.~\onlinecite{Li2020}, we consider an alternative way to define the Pfaffian of a general matrix, using the antisymmetrized matrix through
\footnote{Here we use a different normalization compared to Ref.~\onlinecite{Li2020}.
Therein only the case of certain unitary $A$'s is considered, and an antisymmetric unitary linear combination is constructed.
It can be shown that in all cases considered by Ref.~\onlinecite{Li2020} the result coincides with the definition of the Pfaffian through the normal form.}
\begin{equation}
\widetilde{\pf} (A) = \pf\left(\frac{A - A^T}{2}\right).
\end{equation}

First we show that this definition coincides with substituting a general matrix into the definition of the Pfaffian as a polynomial of the matrix entries
\begin{equation}
\pf (A) = \frac{1}{2^n n!} \sum_{\pi\in S_{2n}} \operatorname{sgn}(\pi)\prod_{i=1}^n a_{\pi(2i-1),\pi(2i)}
\end{equation}
where $\pi$ runs over all permutations of $2n$ elements and $\operatorname{sgn}$ is the signature of the permutation.
Substituting the definition
\begin{align}
&\widetilde{\pf} (A) = \pf\left(\frac{A - A^T}{2}\right) = \nonumber\\
&\frac{1}{2^{2n} n!} \sum_{\pi\in S_{2n}} \operatorname{sgn}(\pi)\prod_{i=1}^n \left(a_{\pi(2i-1),\pi(2i)} - a_{\pi(2i),\pi(2i-1)}\right)=\nonumber\\
&\frac{1}{2^{2n} n!} \sum_{\pi\in S_{2n}} \operatorname{sgn}(\pi)\sum_{\pi'\in S_{\rm pair}}\operatorname{sgn}(\pi')\prod_{i=1}^n a_{\pi'\pi(2i-1),\pi'\pi(2i)}=\nonumber\\
&\frac{1}{2^{2n} n!} \sum_{\pi'\in S_{\rm pair}}\sum_{\pi\in S_{2n}}\operatorname{sgn}(\pi'\pi)\prod_{i=1}^n a_{\pi'\pi(2i-1),\pi'\pi(2i)}=\nonumber\\
&\frac{1}{2^{n} n!} \sum_{\tilde{\pi}\in S_{2n}}\operatorname{sgn}(\tilde{\pi})\prod_{i=1}^n a_{\tilde{\pi}(2i-1),\tilde{\pi}(2i)}.
\end{align}
In the third line we expand the product of differences, and use that the matrix element in the transpose position can be indexed using a permutation which only contains exchanges of adjacent pairs $2i - 1$ and $2i$, with $S_{\rm pair}$ denoting the subgroup of permutations only consisting of such pair exchanges.
In the fourth line we switch the order of the sums and merge the signatures.
In the last line we use that with any fixed $\pi'$, $\pi'\pi = \tilde{\pi}$ runs over $S_{2n}$, hence the sum over $S_{\rm pair}$ can be replaced by $|S_{\rm pair}| = 2^n$.
We emphasize that for general matrix the polynomial expression of the Pfaffian is not identical to the square root of the determinant, this is only true for antisymmetric matrices.

Next we enumerate the properties of $\widetilde{\pf}$, finding that all properties except for the determinant identity \eqref{eqn:pf_square}, the inverse identity \eqref{eqn:pf_inv} and the derivative identity \eqref{eqn:pf_diff} hold for this definition.
We also find that this definition, being a polynomial of the matrix entries, is an everywhere continuous and differentiable function of the matrix.

Finally we establish its relation to our previous generalization of the Pfaffian.
Using the definition \eqref{eqn:gen_pf_def} and the normal form of the antisymmetrized matrix \eqref{eqn:as_nf} we see that the magnitude of $\widetilde{\pf} (A)$ is different from the magnitude of $\pf (A)$ unless $A$ is antisymmetric, but their phases agree, $\arg \widetilde{\pf} (A) = \arg\pf (A)$.
We summarise their relation as
\begin{equation}
\label{eqn:pf_alt_def}
\pf (A) = \sqrt{\frac{\det (A)}{\det\left(\frac{A - A^T}{2}\right)}} \widetilde{\pf} (A),
\end{equation}
where the fraction under the square root is always positive real.
This relation is only valid for conjugate-normal $A$ where $\pf (A)$ is defined, while $\widetilde{\pf} (A)$ is well defined for any matrix.
It is tempting to take \eqref{eqn:pf_alt_def} as a definition to generalize the Pfaffian to all matrices with non-singular antisymmetric part.
The fraction under the square root, however, can have arbitrary complex phase for general matrices, introducing an ambiguity in the branch choice of the square root.
Hence, for non-conjugate-normal matrices the phase of $\widetilde{\pf}$ lacks a connection to the phase of the determinant.

These relations to the the antisymmetrized matrix make it possible to reduce the computation of the Wigner normal form and the generalized Pfaffian to known methods applicable to antisymmetric matrices.
We implemented such an algorithm based on Ref.~\onlinecite{Wimmer2012} in Python, the source code is available on Zenodo, see Ref.~\onlinecite{zenodo}.

\section{Conclusion}
In this note we reviewed two closely related generalizations of the Pfaffian to non-antisymmetric matrices, investigated their relation to each other and to the Wigner normal form of antilinear operators.
The question of whether a generalization of the Pfaffian to non-normal antilinear operators can be defined in a way such that it is the square-root of the determinant is left open.

\begin{acknowledgments}
D.V. is grateful to Anton R. Akhmerov, Isidora Araya Day, \'{A}kos Nagy, Kim P\"{o}yh\"{o}nen, Anastasiia Varentcova, and Michael Wimmer for enlightening discussions and collaboration on related projects.
D. V.  acknowledges funding from NWO VIDI grant 680-47-53, the Swedish Research Council (VR) and the Knut and Alice Wallenberg Foundation.
\end{acknowledgments}

\bibliography{bibliography.bib}

\end{document}